\documentstyle[prl,aps,epsfig,floats] {revtex}
\begin{document}
\draft

\twocolumn[\hsize\textwidth\columnwidth\hsize\csname %
@twocolumnfalse\endcsname

\title{Magnetoresistance of magnetic multilayers in the CPP mode: \\ evidence for
non-local scattering } \author{Didier Bozec, M. A. Howson, and B. J. Hickey}
\address{Department of Physics and Astronomy\\ E. C. Stoner Laboratory, University
of Leeds, Leeds LS2 9JT, United Kingdom} \author{Smadar Shatz  and  Nathan Wiser}
\address{Jack and Pearl Resnick Institute for Advanced Technology\\ Department of
Physics, Bar-Ilan University, Ramat-Gan, Israel} \date{\today} \maketitle

\begin{abstract} We have carried out measurements of the magnetoresistance MR(H)
in the CPP (Current Perpendicular to the Plane) mode for two types of magnetic
multilayers which have different layer ordering. The series resistor model
predicts that CPP MR(H) is {\em independent} of the ordering of the layers.
Nevertheless, the measured MR(H) curves were found to be completely different for
the following two configurations:
\mbox{[}Co(10\AA)/Cu(200\AA)/Co(60\AA)/Cu(200\AA)\mbox{]}$_{\rm N}$ and
\mbox{[}Co(10\AA)/Cu(200\AA)\mbox{]}$_{\rm N}$
\mbox{[}Co(60\AA)/Cu(200\AA)\mbox{]}$_{\rm N}$ showing that the above model is
incorrect.   We have carried out a calculation showing that these results can be
explained quantitatively in terms of the non-local character of the electron
scattering, without the need to invoke spin-flip scattering or a short spin
diffusion length.

\end{abstract}
\pacs{PACS numbers: 75.70.Pa, 75.70.-i, 73.40.-c } ]

Since the discovery a decade ago of the giant magnetoresistance exhibited by
magnetic multilayers, interest in this phenomenon has not abated \cite{1}. Recent
research has focused on the magnetoresistance MR(H) in the CPP mode (current
perpendicular to the plane of the layers) \cite{2,3,4,5,6}. Measurements of MR(H)
are technically more difficult in the CPP mode than in the CIP mode (current in
plane).  However, there are advantages to the MR(H) data in the CPP mode.  For
example, it has been shown \cite{7,8} that experimental values of MR(H) in the
CPP mode can shed light on the spin diffusion length.  Here we present evidence
for the importance of MR(H) measurements in the CPP mode for determining the role
of non-local electron scattering in the giant magnetoresistance (GMR). We shall
show that because of the long electron mean free path, non-local scattering makes
the series resistor model inappropriate.

As is well known, the GMR occurs in magnetic multilayers because the spin-up
electrons and the spin-down electrons have different scattering rates.  If the
electron does not flip its spin upon scattering, then the spin-up and spin-down
electrons constitute two separate currents, with different resistivities, as if
flowing in two parallel wires. In the CPP mode, the resistances of the different
layers add in series \cite{1,7,8}.  Therefore, it would seem that two magnetic
multilayers that differ only in the ordering of the layers would yield identical
results for MR(H) in the CPP mode.

To test this idea, Pratt and co-workers at Michigan State University (Chiang $et$
$al.$ \cite{9}) measured CPP MR(H) for the two configurations [Py/Cu/Co/Cu]$_{\rm
N}$ and [Py/Cu]$_{\rm N}$[Co/Cu]$_{\rm N}$ (denoted as {\lq}interleaved{\rq} and
{\lq}separated{\rq} configurations, respectively), where Py is
Ni$_{84}$Fe$_{16}$.  Although the expectation was that identical MR(H) curves
would be obtained for the interleaved and the separated configurations, these
workers found that the resulting two MR(H) curves were completely different.
Chiang $et$ $al.$\cite{9} attributed their results to the short spin diffusion
length in Py.  They had previously analyzed resistivity data within the framework
of Valet-Fert theory \cite{7,8} and obtained \cite{10} for Py a spin diffusion
length of only 55 \AA, thus implying significant mixing between the spin-up and
spin-down electron currents. Chiang $et$ $al.$ proposed that this spin-flipping
was responsible for the different CPP MR(H) curves they observed for the
separated and interleaved configurations.

We have investigated these ideas by measuring MR(H) for multilayers whose
magnetic layers do $not$ exhibit a short spin diffusion length. For the different
magnetic layers, we used Co of two different thicknesses, since Co is known
\cite{11,12} to have a long spin diffusion length. Measurements were carried out
of CPP MR(H) for [Co(10\AA)/Cu(200\AA)/Co(60\AA)/Cu(200\AA)]$_{\rm N}$ and
[Co(10\AA)/Cu(200\AA)]$_{\rm N}$[Co(60\AA)/Cu(200\AA)]$_{\rm N}$ for \\N = 4, 6,
8. The thickness (200 \AA) of the non-magnetic layers was chosen to be large
enough to ensure complete magnetic decoupling between the ferromagnetic layers.
In spite of the fact that the interleaved and separated configurations differ
only in the ordering of the layers, the measured MR(H) curves were found to be
very different for the two different configurations.  We shall show that these
results can be explained quantitatively in terms of non-local electron
scattering.

The multilayers were grown in our VG-80M MBE facility which has base pressure of
typically 4$\times 10^{-11}$ mbar.  Our CPP measurements used the superconducting
Nb electrode technique, as developed by Pratt {\em et al.} \cite{2}.  The
superconducting equipotential \cite{3,4} ensures that the current is
perpendicular to the layers.  We used a SQUID-based current comparator, working
at 0.1${\%}$ precision to measure changes in the sample resistance of order 10
p$\Omega$.  To avoid driving the Nb normal, the CPP measurements were performed
at 4.2 K in magnetic fields below 3 kOe.  Consistency between the interleaved and
separated samples was enhanced by growing the two configurations during the same
run for each value of N.

The magnetoresistance was measured in the CPP mode for the two configurations:
interleaved and separated. The measured curves for MR(H) are presented for three
values of N in Figs. 1a-1c.  The squares represent the MR(H) data in the
interleaved configuration whereas the circles give the data in the separated
configuration. For each sample, the saturation magnetic field was about 2 kOe.
\begin{figure} \centerline{ \epsfig{figure=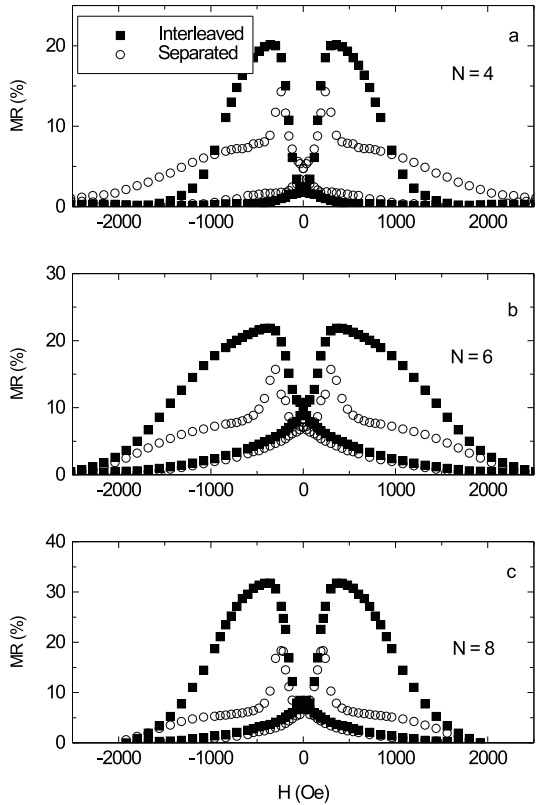,width=7cm}} \vskip 0.5cm
\caption{Magnetic field dependence of the magnetoresistance MR(H) for the
interleaved (squares) and separated (circles) multilayers containing Co (10\AA)
and  Co (60\AA) as the two magnetic metals, for the indicated number of repeats.}
\label{Fig1} \end{figure}

There are several characteristic features of these data, all of which can be
explained in terms of non-local electron scattering. (i) The most important
feature is surely the striking difference between the MR(H) curves for the two
configurations, both in shape and in magnitude. \\(ii) For each N, the maximum
value of MR(H) is larger for the interleaved configuration. (iii) The MR(H) curve
for the interleaved configuration exhibits a single peak, whereas for the
separated configuration, MR(H) is the superposition of two peaks, with the second
being much broader and less delineated than the first. Another interesting
feature of the data, not displayed in Fig. 1, is that the saturated resistance
itself is always greater for the interleaved configuration.

To ensure that the differing results for MR(H) for the two configurations are not
due to differences in their magnetic properties, the magnetization as a function
of field was measured for each sample.  We found that the two configurations
yield the same magnetization. This confirms that the magnetic layers are
uncoupled and become magnetized independently. At low fields, the magnetization
curves are dominated by the contribution of the thicker Co layers.  After the
thicker Co layers reach saturation, the magnetization continues to increase as
the thinner Co layers approach saturation. The magnitudes of the saturation
fields for the two thicknesses of Co layers correspond closely to the saturation
fields of MR(H).

Kinetic theory arguments show that the electron mean free path is far longer than
the thicknesses of the magnetic layers ({10 \AA} and 60 \AA). Therefore, the
potential "felt" by the electron is the combined potential of a neighboring pair
of magnetic layers. This may be termed "non-local" electron scattering in the
sense that one cannot speak of the resistivity of a $single$ Co layer. Rather,
the resistivity is determined by a property of $pairs$ of neighboring layers.
Gittleman $et$ $al.$ \cite{13} have shown that for such a case, the contribution
of the spin-direction-dependent resistivity depends on the cosine of the angle
$\theta_{ij}$ between the moments of neighboring magnetic layers, {\em i} and
{\em j}.  This is the key to understanding the data.

Because the mean free path is larger than the layer thicknesses, it is necessary
to carry out a full band structure calculation to calculate properly the
resistivity and magnetoresistance. However, one can understand the basic physics
with a simple phenomenological model.

For the interleaved configuration, the neighboring magnetic layers are different,
and hence the maximum angle $\theta_{ij}$ is large, whereas for the separated
configuration, the neighboring magnetic layers are the same (except for one
boundary layer), and hence the maximum angle $\theta_{ij}$ is small. Therefore,
there is no reason to expect MR(H) to be the same for the two configurations.
This explains the first feature of the data mentioned above.

From the above considerations, it also immediately follows that MR(H) will be
larger for interleaved multilayers than for separated multilayers, because the
angle $\theta_{ij}$ is larger for the former configuration.  This explains the
second feature of the data mentioned above. This has been confirmed by
measurements of the GMR as a function of the number of bilayers. A
Fuchs-Sondheimer analysis of these data shows that the mean free path in
sputtered \cite{14} and MBE \cite{15} samples is about {500 \AA} and {700 \AA},
respectively.

For the interleaved configuration, there is only $one$ angle $\theta_{ij}$ that
is relevant, namely, the angle between the moments of the different ({10\AA} and
{60\AA}) neighboring magnetic layers. Therefore, there will be only $one$ peak,
as the angle $\theta_{ij}$ becomes progressively larger, passes through a maximum
at the saturation field of the Co ({60\AA}) layer and then becomes smaller as the
Co ({10\AA}) layer also saturates. By contrast, for the separated configuration,
there are $two$ angles $\theta_{ij}$ that are relevant, namely, the angle between
neighboring moments for each set of layers (the {10 \AA} set and the {60 \AA}
set). As each angle $\theta_{ij}$ passes through its maximum, a peak will be
obtained for MR(H), leading to two overlapping peaks, with each maximum occurring
at a different value of the magnetic field, corresponding roughly to the coercive
field of each type of magnetic layer.  This explains the third feature of the
data mentioned above.

These ideas can be made quantitative. If the spin diffusion length is very long,
it is known \cite{1} that a simple expression is obtained for MR(H). According to
the phenomenological theory of Wiser \cite{17}, for the geometry under
consideration here and assuming a very long spin diffusion length, the
magnetoresistance due to an $ij$-pair of neighboring magnetic layers is:
\begin{equation} MR_{ij}(H)=c_{ij}(1-\cos\theta_{ij}(H))^2
\end{equation}

The spin diffusion length of Co has been measured yielding values of {450 \AA}
\cite{11,12} and {1000 \AA} \cite{9}. These values is very much larger than the
thickness of the Co layers, and so one may safely employ the expression for MR(H)
given in (1).

For our samples, there are three parameters $c_{ij}$ corresponding to the three
different types of neighboring pairs of magnetic layers: i = j = 1; i = j = 2; i
= 1, j = 2, where 1 refers to Co ({60\AA}) layers and 2 refers to Co ({10\AA})
layers.  The interleaved configuration contains only type i = 1, j = 2 neighbors,
whereas the separated configuration contains all three types.  For a sample
containing N repeats, the separated configuration consists of N-1 pairs of type i
= j = 1 neighbors, followed by one pair of type i = 1, j = 2 neighbors (the
boundary layer), followed by N-1 pairs of type i = j = 2 neighbors.

First consider the interleaved configuration.  The saturation magnetic field
$H_{s1}$ of the thicker Co layers is smaller than $H_{s2}$ of the thinner Co
layers.  Thus, as the magnetic field is increased, the angle $\theta_{1,2}$
increases, since the thicker Co layers are reversing their direction of
magnetization faster than the thinner Co layers. According to Eq. (1), increasing
the angle $\theta_{1,2}$ implies an increase in MR(H).  When the magnetic field
reaches $H_{s1}$, the angle $\theta_{1,2}$ reaches its maximum value, and begins
to decrease as the thinner Co layers continue to reverse their direction of
magnetization while the thicker Co layers have already reached saturation.
According to Eq. (1), decreasing $\theta_{1,2}$ leads to a decrease in MR.
Finally, when the field reaches $H_{s2}$, the angle $\theta_{1,2}$ is again zero,
and MR vanishes.  Thus, we expect - and find - a single peak for MR(H) for the
interleaved configuration.

The field dependence of $\theta_{ij}$ is determined as follows.  The
magnetization increases linearly with field (except near saturation, where it
increases more slowly).  Since the magnetization is proportional to the cosine of
the angle between the magnetic moment and the field, it follows that
cos$\theta_{i}$ and cos$\theta_{j}$ are each linear in the field. Equation (1)
contains cos$\theta_{ij}$ = cos($\theta_{i}$ - $\theta_{j}$). Expanding the
cosine readily gives the required field dependence.

The calculated results \cite{17} for the interleaved configuration are given by
the curves in Fig. 2. For each value of N, the parameter $c_{1,2}$ was determined
by fitting to the MR(H) data. The agreement between the calculated curves and the
data is evident from the figure.
\begin{figure} \centerline{ \epsfig{figure=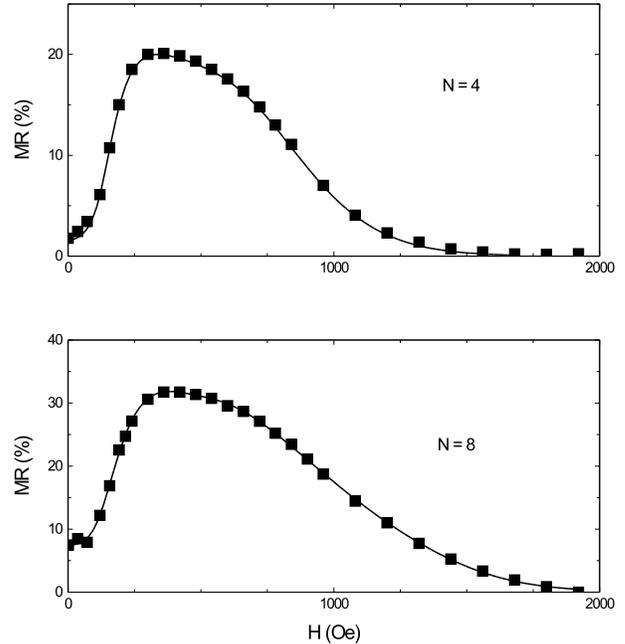,width=8cm}} \vskip 0.5cm
\caption{Comparison between the calculated curves (solid lines) and the data
points (squares) for MR(H) for the interleaved configuration for the indicated
number of repeats.} \label{Fig2}
\end{figure}

We now consider the separated configuration.  If the Co layers were ideal
single-domain structures, then the magnetic moment of each Co layer would react
identically to the magnetic field and the angles $\theta_{1,1}$ and
$\theta_{2,2}$ would both be zero at all fields.  However, because of the
presence of domains and of structural imperfections in the Co layers, each layer
reverses its magnetization at a somewhat different rate.  As a result, the angles
$\theta_{1,1}$ and $\theta_{2,2}$ become non-zero as the field is increased, pass
through a maximum at the coercive field, and then decrease to zero as saturation
is approached.

We assumed a simple parabolic form for each of the two angles. The maximum value
of each parabola, $\theta_{max,1,1}$ and $\theta_{max,2,2}$, cannot be determined
by fitting to the data for the following reason.  Because these angles are small,
Eq. (1) can be expanded to yield
\begin{equation} MR_{ii}= c_{ii}(\slantfrac{1}{2}\theta_{ii}^2)^2\propto
c_{ii}(\theta_{max,ii})^4
\end{equation} and this $combination$ of $c_{ii}$ and $\theta_{max,ii}$ serves as
a $single$ fitting parameter.  Nevertheless, some numerical tests we have carried
out suggest that both $\theta_{max,1,1}$ and $\theta_{max,2,2}$ lie in the range
of $15^{\circ}-30^{\circ}$.  This value is, of course, much smaller than the
maximum value of the angle $\theta_{1,2}$. This explains why MR(H) is larger for
the interleaved configuration than for the separated configuration.
\begin{figure} \centerline{ \epsfig{figure=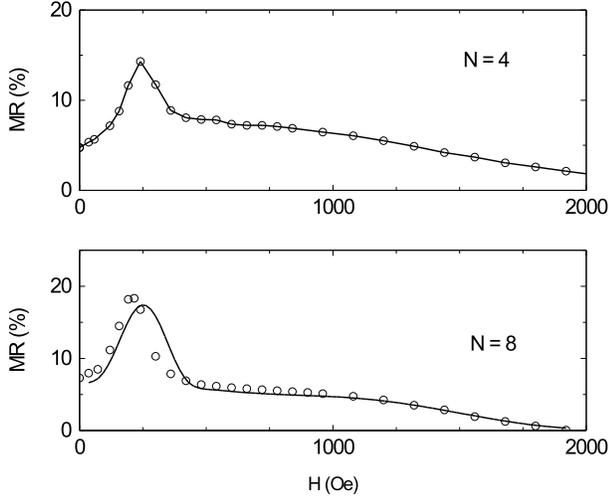,width=8cm}} \vskip 0.5cm
\caption{Comparison between the calculated curves (solid lines) and the data
points (circles) for MR(H) for the separated configuration for the indicated
number of repeats.} \label{Fig3} \end{figure}

The calculated results \cite{17} for the separated configuration are given by the
curves in Fig. 3.  For each value of N, the three parameters
$c_{1,1}(\theta_{max,1,1})^4$, $c_{2,2}(\theta_{max,2,2})^4$, and $c_{1,2}$ were
determined by fitting to the MR(H) data.  The agreement between the calculated
curves and the data is evident from the figure.

To confirm that MR(H) for the separated configuration contains the contributions
of {[}Co(10\AA)/Cu(200\AA){]}$_{\rm N}$ and of {[}Co(60\AA)/Cu(200\AA){]}$_{\rm
N}$, we also measured MR(H) for a multilayer containing only
{[}Co(10\AA)/Cu(200\AA){]}$_{\rm N}$ and for another multilayer containing only
{[}Co(60\AA)/Cu(200\AA){]}$_{\rm N}$. For each of these two multilayers, MR(H)
consists of a single peak, located at the same magnetic field as one of the two
peaks in the separated configuration. Thus, the two peaks observed for the
separated configuration do indeed correspond to the two individual peaks.

In conclusion, we have shown that the principal features of the MR(H) data can be
explained quantitatively, for both the interleaved and the separated
configurations, by invoking non-local electron scattering.

It is a pleasant duty to acknowledge that this research was supported by grants
from the UK-Israel Science and Technology Research Fund and the UK-EPSRC. We
appreciate discussions with C. H. Marrows and A. Carrington. D. Bozec thanks the
University of Leeds for financial support.

\end{document}